\documentclass{ws-mpla}

\newcommand{\be}{\begin{equation}}
\newcommand{\ee}{\end{equation}}

\newcommand{\no}{\noindent}
\newcommand{\ba}{\begin{eqnarray}}
\newcommand{\ea}{\end{eqnarray}}
\newcommand*{\pd}{\partial}
\newcommand*{\pdm}{\pd_{\mu}}
\newcommand*{\pdn}{\pd_{\nu}}

\newcommand*{\bea}{\begin{eqnarray}}
\newcommand*{\eea}{\end{eqnarray}}
\newcommand*{\pref}[1]{(\ref{#1})}

\newcommand*{\prefr}[2]{(\ref{#1}-\ref{#2})} 

\begin{document}

\markboth{Axel Maas}{Landau gauge Yang--Mills theory at finite temperature}

%
\catchline{}{}{}{}{}
%

\title{Gluons at finite temperature in Landau gauge Yang--Mills theory}

\author{\footnotesize Axel Maas}

\address{Gesellschaft f\"ur Schwerionenforschung mbH, Planckstra{\ss}e 1,\\ D-64291 Darmstadt, Germany\\Axel.Maas@Physik.TU-Darmstadt.de}

\maketitle

\pub{Received (Day Month Year)}{Revised (Day Month Year)}

\begin{abstract}

The infrared behavior of Yang-Mills theory at finite temperature provides access to the role of confinement. In this review recent results on this topic from lattice calculations and especially Dyson--Schwinger studies are discussed. These indicate persistence of a residual confinement even in the high--temperature phase. The confinement mechanism is very similar to the one in the vacuum for the chromomagnetic sector. In the chromoelectric sector screening occurs at the soft scale $g^2T$, although not leading to a perturbative behavior.

\end{abstract}

\ccode{PACS Nos.: 11.10.Wx 11.15.-q 11.15.Tk 12.38.-t 12.38.Aw 12.38.Lg 12.38.Mh 14.70.-e 14.70.Dj 25.75.Nq}

\section{Introduction}

One of the most intriguing properties and challenging problems in QCD is confinement. In recent years, the understanding of gluon confinement in Landau gauge has progressed\cite{Alkofer:2000wg,Zwanziger:2003cf}, in a combined effort of methods using Dyson--Schwinger equations (DSEs)\cite{vonSmekal:1997is,Fischer:2002hn}, renormalization group (RG) methods\cite{Gies:2002af,Pawlowski:2003hq} and in a multitude of lattice calculations (for the most recent results, see {\it e.g.} Bowman {\it et al.}\cite{Bowman:2004jm} and Oliveira {\it et al.}\cite{Oliveira:2004gy}). The question of quark confinement, on the other hand, is not yet at the same stage\cite{Alkofer:2003jk}. However, significant evidence exists that gluon confinement is not affected by a small (physical) number of light quarks\cite{Bowman:2004jm,Fischer:2003rp}. Furthermore the non--perturbative features of QCD are most probably generated in the gauge sector. It is therefore reasonable to use the knowledge on gluon confinement also in applications, like the investigation of Yang-Mills theory at finite temperature.

Several decades ago it was argued\cite{Cabibbo:1975ig} that QCD and Yang--Mills theory undergo a deconfining phase transition at sufficiently high temperatures. This prediction has been investigated in many experiments\cite{Andronic:2004tx} and in lattice calculations\cite{Karsch:2003jg}. One of the striking expectations was that the high--temperature phase should consist of weakly interacting quasi--particles. However, although thermodynamic bulk properties seem to logarithmically approach this behavior\cite{Karsch:2003jg,Karsch:2001cy}, the microscopic properties, especially in the chromomagnetic sector, do not seem to behave as expected\cite{Rischke:2003mt}. Also there is not yet a theoretical explanation which can consistently explain all experimental observations. Many attempts to reconcile this problem, {\it e.g.\ } hard thermal loops (HTL)\cite{Pisarski:1988vd,Weldon:aq,Blaizot:2001nr}, have been made, but did not succeed. This was to some extent anticipated, as the infinite-temperature limit of Yang--Mills theory is equivalent to a 3--dimensional (3d) Yang--Mills theory coupled to a massive Higgs field\cite{Appelquist:vg,Kajantie:1995dw}. 3d Yang--Mills theory is confining\cite{Feynman:1981ss}, and thus non--perturbative interactions have to be expected.

Here, recent results on this subject are reviewed, especially from DSEs\cite{Gruter:2004kd,Maas:2004se,Maas:2005hs}. With a direct handle on gluon confinement available, it is possible to trace the fate of confinement along the complete temperature axis. As expected gluon confinement prevails below the phase transition\cite{Gruter:2004kd}. In the high--temperature phase, part of the gluon spectrum is still confined\cite{Maas:2005hs}. This confinement persists, as anticipated, in the infinite--temperature limit\cite{Maas:2004se}. These results agree with findings in lattice calculations\cite{Cucchieri:2003di,Cucchieri:2001tw}.

This review is organized as follows. In section \ref{sym} basic properties of Yang--Mills theory are collected, including signals of confinement. Aspects of the formulation at finite temperature are presented in section \ref{sdse}, together with a discussion of the DSEs. In section \ref{sresults} the results are given. These are discussed in section \ref{sdiscuss}, which includes the impact on the phase diagram and the possible influence of quarks. This section also concludes the review.

\section{Aspects of strong Yang--Mills theory}\label{sym}

The theory studied is an equilibrium Yang--Mills theory, governed in the Matsubara formalism by the Euclidean Lagrangian\cite{Alkofer:2000wg,Kapusta:tk}
\bea
\mathcal{L}&=&\frac{1}{4}F_{\mu\nu}^aF_{\mu\nu}^a+\bar c^a \pdm D_\mu^{ab} c^b\label{lagrange}\\
F^a_{\mu\nu}&=&\pdm A_\nu^a-\pdn A_\mu^a-gf^{abc}A_\mu^bA_\nu^c\nonumber\\
D_\mu^{ab}&=&\delta^{ab}\pdm+gf^{abc}A_\mu^c\nonumber~.
\eea
\no Here $F_{\mu\nu}^a$ denotes the field strength tensor, $D_\mu^{ab}$ the covariant derivative, $g$ the gauge coupling, and $f^{abc}$ the structure constants of the gauge group. $A_\mu^a$ is the gluon field and $\bar c^a$ and $c^a$  are the Faddeev--Popov ghost fields, describing part of the intermediate states of the gluon field\footnote{The hermiticity assignment is valid in Landau gauge, although it is not the conventional one\cite{Alkofer:2000wg}.}. The gauge chosen is the Landau gauge for technical reasons, most importantly because of the non--renormalization of the ghost--gluon vertex\cite{Taylor:ff}. At finite temperature no comparable set of investigations in other gauges have yet been completed. There are investigations in the vacuum in Coulomb gauge using DSEs\cite{Feuchter:2004gb} and lattice methods\cite{Langfeld:2004qs}, which indicate a gluon confinement mechanism equivalent to that in Landau gauge. Investigations in other gauges do not yet lead to a satisfactory understanding of gluon confinement even in the vacuum, as the assumptions made turned out to be too restrictive\cite{Alkofer:2000wg,Alkofer:2003jr}. Thus it is still unclear how the observations collected here translate to other gauges.

A central issue is the fate of confinement and what ``deconfinement'' implies. Therefore adequate criteria are needed to test confinement. These are partly encoded in the pertinent 2--point functions. The relevant properties will be only listed here. A brief review of them can be found elsewhere\cite{Schleifenbaum:2004mm}.

Empirically, if the spectral function of a particle is not positive semi--definite, no K\"all{\'e}n--Lehmann representation exists. It is then not part of the physical spectrum and thus confined\cite{Oehme:bj}. A violation of positivity occurs, if the corresponding propagator $D$ vanishes at zero momentum
\be 
\lim_{p^2 \to 0}D(p^2)=0.\label{Oehme} 
\ee
\no This is a sufficient, but not a necessary condition. Intuitively, for a massless particle condition \pref{Oehme} yields a vanishing would--be on--shell propagator. Thus the corresponding particle does not propagate and is confined.

The behavior in eq.\ \pref{Oehme} is also predicted by two confinement scenarios. The one of Kugo and Ojima\cite{Kugo:gm} puts forward the idea that all colored objects form BRST quartets and are thus unphysical. One precondition for this is an unbroken global color charge. In the Landau gauge, this condition can be cast into\cite{Kugo:1995km}
\be 
\lim_{p^2 \to 0} p^2 D_G(p^2)\to\infty,\label{Kugo} 
\ee
\no where $D_G$ is the propagator of the Faddeev--Popov ghost. Such an infrared divergence relates to long--ranged spatial correlations, stronger than those induced by a Coulomb force, since the divergence in momentum space is stronger than that of a massless particle. In this sense, ghosts mediate confinement.

In the Gribov--Zwanziger scenario\cite{Zwanziger:2003cf,Gribov:1977wm,Zwanziger:2001kw}, entropy arguments suggest the dominance of field configurations close to or on the Gribov horizon in field configuration space. The Gribov horizon in Landau gauge is characterized by zeros of the inverse ghost propagator, predicting eq.\ \pref{Kugo}. For an infrared constant ghost--gluon vertex, which is supported by lattice simulations\cite{Cucchieri:2004sq} and calculations based on DSEs\cite{Schleifenbaum:2004id,Alkofer:2004it}, eq.\ \pref{Oehme} follows for the gluon propagator\cite{Zwanziger:2003cf,vonSmekal:1997is}. A further consequence of this scenario is that the truncation introduced below becomes exact in the infrared.

\section{Finite temperature and Dyson-Schwinger equations}\label{sdse}

The basic quantity encoding the content of a theory is the infinite set of Green's functions. Their knowledge allows to calculate any thermodynamic quantity, as they completely determine the partition function.

In practice, however, only a finite number of Green's functions can be determined within approximation schemes. These can be used to estimate thermodynamic quantities using e.g.\ the Luttinger-Ward/Cornwall-Jackiw-Tomboulis formalism\cite{Luttinger:1960ua}. The thermodynamic potential can alternatively be determined by lattice calculations. However, in most cases its microscopic origin remains elusive. Furthermore, as infrared singularities are anticipated according to condition \pref{Kugo}, a complementary continuum formulation, not plagued by finite volume effects, is desirable.

The Green's functions can be determined by the DSEs, which are the quantum analog of the classical equations of motion\cite{Alkofer:2000wg,Dyson:1949ha}. Finite temperature can be introduced into the equations using the Matsubara formalism\cite{Kapusta:tk}, which compactifies the Euclidean time dimension to a length of $1/T$. Thus, the zero component of the momentum becomes discrete with $p_0=2\pi n T$ for bosons, where $n$ is an integer including zero\footnote{Note that this also applies to ghosts despite their Grassmannian nature, as they live on the periodic representation of the gauge group\cite{Bernard:1974bq}.}. Furthermore, Lorentz invariance is no longer manifest\cite{Weldon:aq}. Therefore, the propagators of the ghost and gluon can be described as\cite{Kapusta:tk}
\bea
D_G(p)&=&-\frac{G(p_0^2,\vec p^2)}{p^2}\label{ghostDressing}\\
D_{\mu\nu}(p)&=&P_{T\mu\nu}(p)\frac{Z(p_0^2,\vec p^2)}{p^2}+P_{L\mu\nu}(p)\frac{H(p_0^2,\vec p^2)}{p^2} \; .\label{gluonDressing}
\eea
\no Here $P_L$ and $P_T$ are projectors longitudinal and transverse to the heat bath, respectively. Both are 4d--transverse, as required by Slavnov--Taylor identities\cite{Kapusta:tk,Bohm:yx}. The soft mode ($p_0=0$) of $P_T$ consists only of space-space (chromomagnetic) components while the one of $P_L$ only contains time-time (chromoelectric) components. $G$, $Z$ and $H$ are scalar, dimensionless dressing functions, fully specifying the propagators. At zero temperature, $Z$ and $H$ coincide.

\begin{figure*}
\begin{center}
\epsfig{file=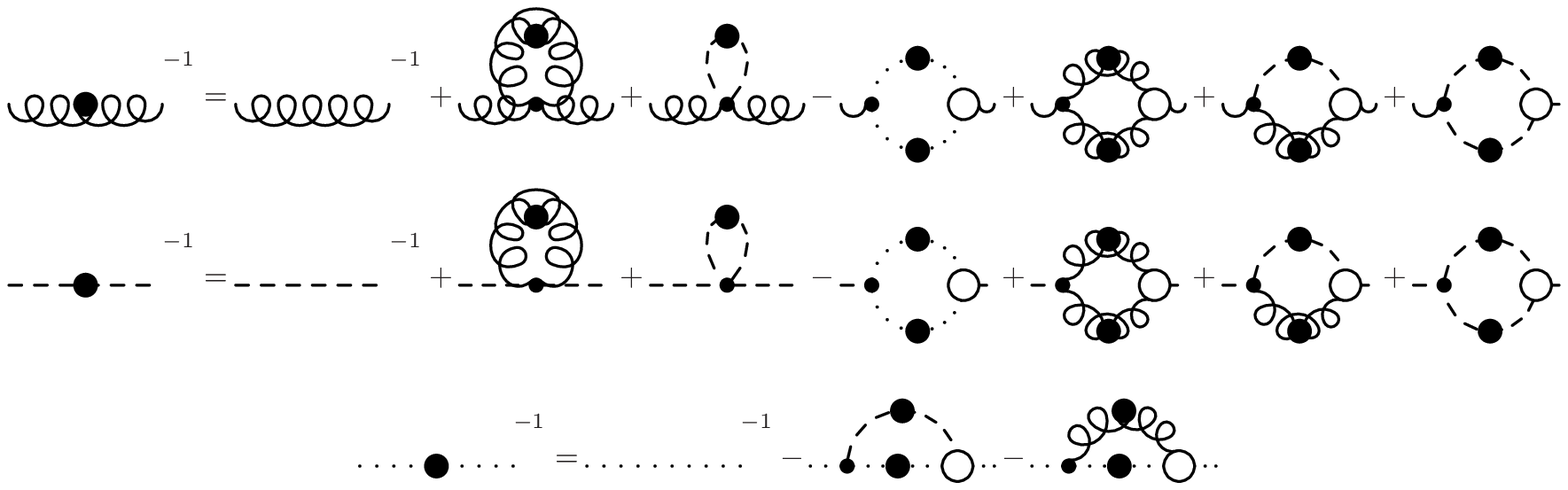,width=0.9\linewidth}
\end{center}
\caption{The truncated Dyson--Schwinger equations at finite temperature. The dotted lines denote ghosts, the dashed lines longitudinal gluons and the wiggly lines transverse gluons. Lines with a full dot represent self--consistent propagators and small dots indicate bare vertices. The open circled vertices are full and must be constructed in a given truncation scheme. A bare ghost--gluon vertex and slightly modified bare gluon vertices have been used here. Note that for soft modes the ghost--longitudinal and the 3--point coupling of three longitudinal and of one longitudinal and two transverse gluons vanish.}\label{figsysft}
\end{figure*}

The DSEs form an infinite set of coupled, non--linear integral equations for the Green's functions. By expanding the Green's functions in powers of $g$ ordinary perturbation theory is recovered. If any progress is to be made beyond resummed perturbation theory, it is necessary to solve these equations self-consistently. This is in general not possible. It is therefore necessary to truncate the system to make it tractable. Truncations are in general a--priori not justifiable. It is therefore necessary to check their validity a--posteriori. The truncation scheme used for the results presented here\cite{vonSmekal:1997is,Gruter:2004kd,Maas:2004se,Maas:2005hs} has been validated in various systematic ways\cite{Fischer:2003rp,Maas:2004se,Schleifenbaum:2004id,Fischer:2003zc,Maas:2005rf,Lerche:2002ep,Alkofer:2004it,Watson:phd,Bloch:2003yu,Atkinson:1998zc}. Some of the assumptions have been tested using lattice calculations\cite{Cucchieri:2004sq,Boucaud:1998xi,Cucchieri:1997ns}. In addition, the results are in very good agreement both qualitative and quantitative with available lattice\cite{Bowman:2004jm,Oliveira:2004gy,Cucchieri:2003di,Cucchieri:2001tw} and RG calculations\cite{Gies:2002af,Pawlowski:2003hq}. The truncated equations are graphically represented in figure \ref{figsysft}.

A serious obstacle in any non--perturbative calculation for gauge theories is the Gribov problem\cite{Gribov:1977wm,Singer:dk}, as a local gauge condition is not able to fully specify the gauge. However, recently is has been argued\cite{Zwanziger:2003cf} that it is sufficient to stay within the first Gribov horizon to evade this problem. This is achieved by requiring the dressing functions $G$, $Z$, and $H$ to be positive semi--definite. The influence of Gribov copies has been studied on the lattice and no contradiction to this assumption has been found\cite{Cucchieri:1997ns}.

A further interesting property is obtained in the limit of infinite temperature. This is equivalent to the static limit of the theory, as the extension of the time direction shrinks to zero. Thus it can be described by a 3d theory\cite{Appelquist:vg}. The corresponding 3d Lagrangian can be obtained as an effective field theory\cite{Kajantie:1995dw}. In this course the original $A_0$ component of the gluon field becomes an additional adjoint Higgs field, thus preserving the number of degrees of freedom, as a 3d gluon field has only one transverse polarization. The constants that appear in the 3d theory, like the Higgs mass, can be obtained by matching with the original theory\cite{Kajantie:1995dw,Maas:2004se,Maas:2005hs}. The most important quantity in this context is the (now dimensionful) 3d gauge coupling $g_3$. Within the truncation scheme employed here\cite{Maas:2005hs}, it is given by $g_3^2=g^2T$. The value of the running coupling $g$ depends on the renormalization scale $\mu$ that can be selected at will. Usually $\mu$ is chosen to be proportional to $T$, yielding at high temperatures a logarithmically decreasing coupling due to asymptotic freedom. However, to obtain a well--defined infinite--temperature limit with a finite 3d gauge coupling $g_3$, it is here necessary to fix the value of $g^2T$ instead\cite{Maas:2004se,Maas:2005hs}. Thus the running coupling decreases polynomially with temperature.

\section{Results}\label{sresults}

\subsection{Asymptotics}

For asymptotically small and large (Euclidean) momenta, the truncated DSEs can be solved analytically. In the ultraviolet this is due to asymptotic freedom. Thus at momenta large compared to $\Lambda_{\mathrm{QCD}}$ and $T$, the same logarithmic running of the dressing functions occurs as in the vacuum\cite{Bohm:yx,Fischer:2003zc}. $Z$ and $H$ coincide in this regime\footnote{At sufficiently small temperatures this is already the case in the non--perturbative domain.}. Using an appropriate three--gluon vertex, resummed perturbation theory to one--loop order is also recovered in the self-consistent solutions of the DSEs\cite{Fischer:2003zc}.

This behavior persists at all temperatures. At sufficiently large temperatures, $T\gg\Lambda_{\mathrm{QCD}}$, the intermediate momentum regime $\Lambda_{\mathrm{QCD}}<p<T$, becomes indistinguishable from the perturbation theory of a 3d theory of the soft modes alone. This is due to the decoupling of the hard modes, which have an effective mass of at least $2\pi T$. This is a hard scale. Thus the hard modes are essentially perturbative and the Appelquist-Carrazone theorem applies. When the infinite--temperature limit is taken, this affects the full momentum range, and the effective 3d theory is obtained\cite{Appelquist:vg}. This behavior is explicitly seen in the DSE results\cite{Maas:2004se,Maas:2005rf}.

In the infrared, the truncated system of DSEs can be solved analytically and self-consistently\cite{vonSmekal:1997is,Maas:2004se,Zwanziger:2001kw,Lerche:2002ep} by power--law ans\"atze for the soft modes,
\bea
G(p)&=&A_g(p^2)^{-\kappa}\label{irg}\\
Z(p)&=&A_z(p^2)^{-t}\\
H(p)&=&A_h(p^2)^{-l}\label{irh}.
\eea
\no The exponents are related\cite{vonSmekal:1997is}. In the vacuum $-2\kappa=t$ and $t=l$, and it can be shown that $\kappa>0$ quite generally\cite{Watson:2001yv}. However, the exact value of $\kappa$ depends on the truncation\cite{Lerche:2002ep}, but strong arguments exist for $\kappa>1/2$. Therefore conditions \pref{Oehme} and \pref{Kugo} are satisfied and gluons are confined in the vacuum. The best value to date, in agreement between DSE\cite{Zwanziger:2001kw,Lerche:2002ep} and RG methods\cite{Pawlowski:2003hq}, is $\kappa\approx 0.595$, a value which is compatible with latest lattice results\cite{Oliveira:2004gy}, suggesting $\kappa\approx 0.53$. Numerical results from DSE calculations at non--zero temperatures below the phase--transition find no change of the exponents\cite{Gruter:2004kd}.

Interestingly enough, the terms stemming from the gauge--fixing part of the Lagrangian \pref{lagrange}, {\it i.e.}\ only diagrams on the r.\ h.\ s.\ in figure \ref{figsysft} with at least one ghost line, dominate in the infrared. This agrees with the Gribov--Zwanziger scenario and thus a Yang--Mills theory would be a topological field theory of Schwarz type in the infrared. Therefore no propagating modes exist and confinement is manifest.

The situation in the high--temperature phase is different. The value of the exponents are again independent of the temperature and can most directly be calculated in the dimensionally reduced theory\cite{Maas:2004se,Zwanziger:2001kw,Maas:2005rf}. In this case $-2\kappa=t+1/2$. Two solutions are found with $\kappa=1/2$ and $\kappa\approx 0.39$, both satisfying conditions \pref{Oehme} and \pref{Kugo}. The phase is therefore strongly interacting and confining, even in the limit of infinite temperature, as was anticipated\cite{Zahed:1999tg}. Considerations\cite{Maas:2004se,Zwanziger:2001kw} concerning smoothness in dimensionality and indications from the thermodynamic potential favor the value $\kappa\approx 0.39$ as do lattice results\cite{Cucchieri:2003di}. Again only the gauge--fixing part contributes in the transverse sector. However, in the longitudinal sector the situation is different. Here $t\neq l$ due to interaction of the soft modes with the hard modes, leading to the generation of a screening mass\cite{Maas:2005hs,Maas:2005rf}. A massive particle has an infrared constant propagator, requiring $l=-1$. This result also agrees with lattice calculations\cite{Cucchieri:2001tw}. Thus, longitudinal modes in contrast to transverse modes are not confined according to condition \pref{Oehme}. Therefore the high--temperature phase is qualitatively different. Confinement in the longitudinal sector is discussed further in section \ref{ssconfinement}.

The hard modes, on the other hand, are restricted to $p^2=p_0^2+\vec p^2>p_0^2$. Infrared is therefore measured with respect to $\vec p^2$. In this sense, the hard modes exhibit in both phases the behavior of massive particles, {\it i.e.} $\kappa_h=t_h=l_h=-1$ as expected. At sufficiently low temperatures non--perturbative effects still contribute, and the hard modes cannot be described accurately by perturbation theory alone.

An interesting feature of the results \prefr{irg}{irh} is a qualitative independence of the gauge group: For any semi--simple compact Lie--group, such as the physical SU(3), the results are qualitatively the same. This is most pronounced in the infinite--temperature limit, where the results, after proper rescaling, are also quantitatively independent of the gauge group. This property hinges on the validity of the approximations made, as here the gauge group only enters through its adjoint Casimir. This factor drops out (as does the gauge coupling) in the determination of the infrared exponents. It is as yet unclear whether this property persist beyond the present truncation. Therefore, in the following only the gauge group SU(3) will be considered, except for the comparison with lattice data, using SU(2) where necessary. If, however, the truncation made here is indeed exact in the infrared, as is suggested by the Gribov--Zwanziger scenario, all such gauge groups have a qualitative unique infrared limit. This also implies that the mechanism of gluon confinement and the phase structure in Landau gauge is independent of the gauge group, which is an interesting observation to be investigated further.

\subsection{Full solutions}

The connection between the analytical infrared and ultraviolet regime has been obtained by lattice calculations as well as by self--consistently solving the DSEs numerically. There are various techniques for the latter case. In the low--temperature regime a torus discretization has been used\cite{Fischer:2003rp,Gruter:2004kd,Fischer:2005ui}. In the high--temperature regime a continuum method was employed\cite{Maas:2005rf,Maas:2005xh}. In the vacuum both have been used\cite{Fischer:2002hn,Atkinson:1998zc,Hauck:1996sm}.

\begin{figure*}
\epsfig{file=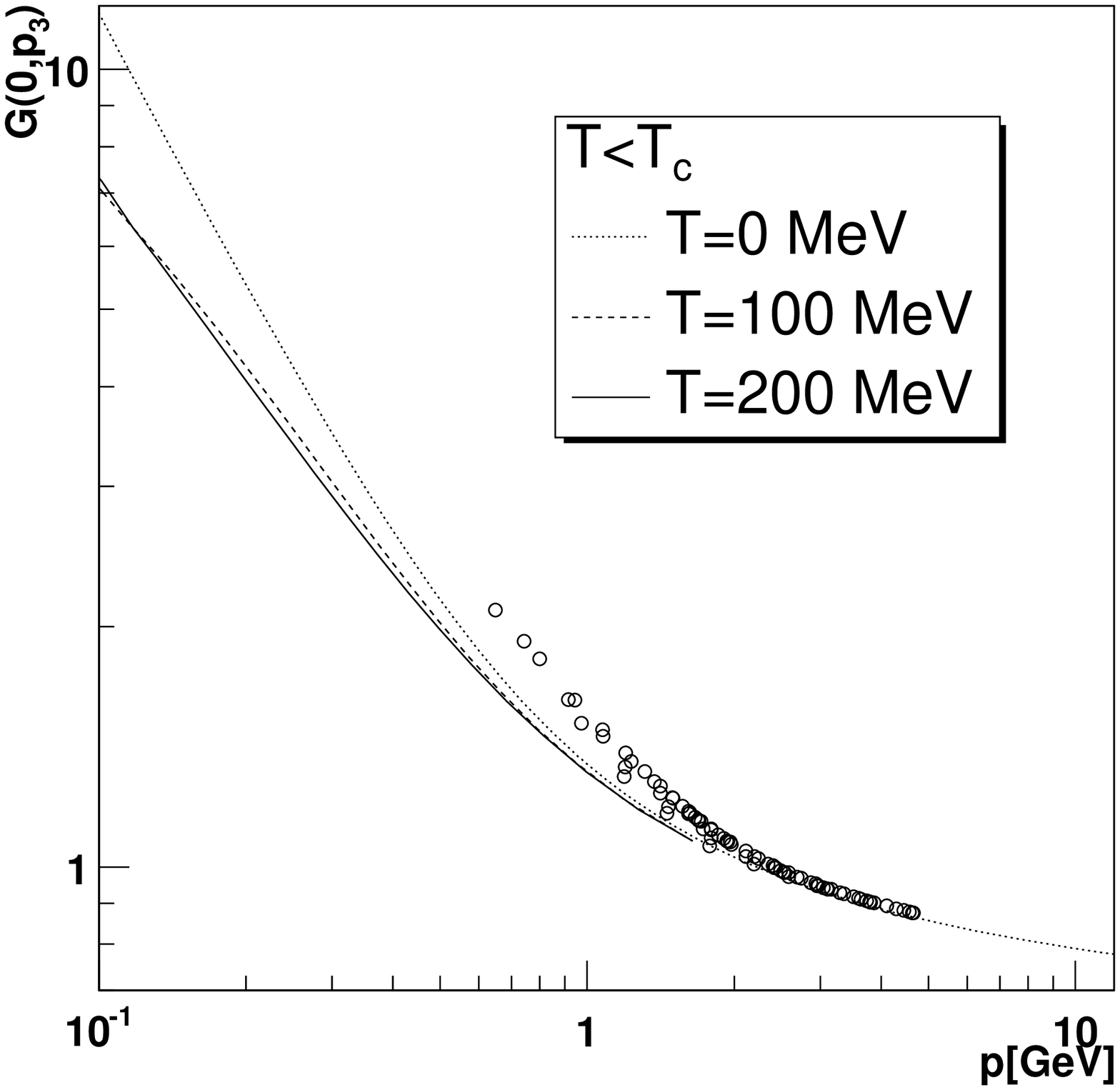, width=0.5\linewidth}\epsfig{file=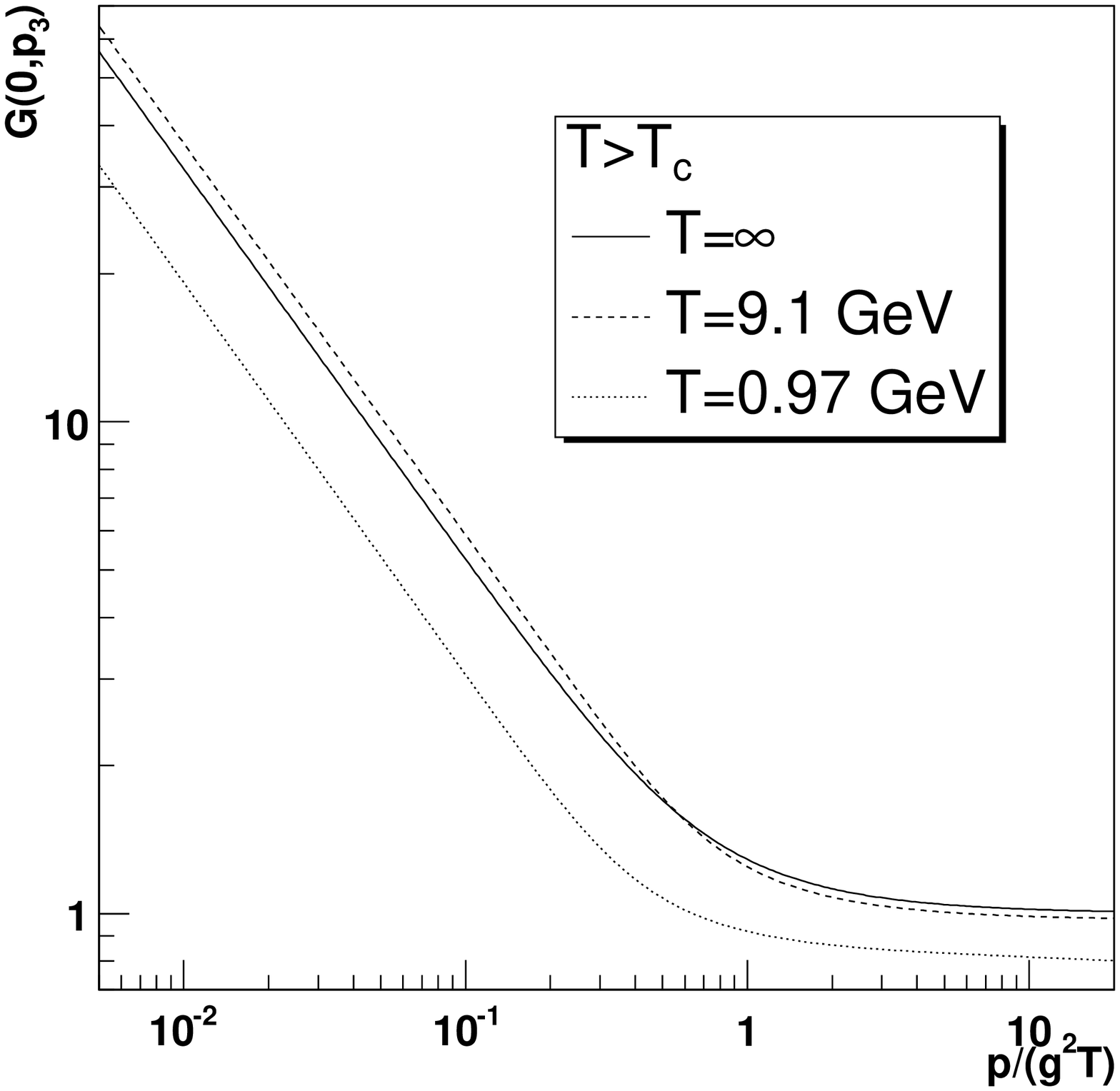, width=0.5\linewidth}
\caption{The ghost dressing function is shown in the vacuum\protect\cite{Fischer:2002hn} and at temperatures below the phase transition\protect\cite{Gruter:2004kd} in the left panel and above the phase transition\protect\cite{Maas:2004se,Maas:2005hs} in the right panel. The lattice data in the left panel are at zero temperature\protect\cite{Langfeld:2002dd}. The three momentum is signified by $p_3=\left|\vec p\right|$.}\label{figg}
\end{figure*}

\begin{figure*}
\epsfig{file=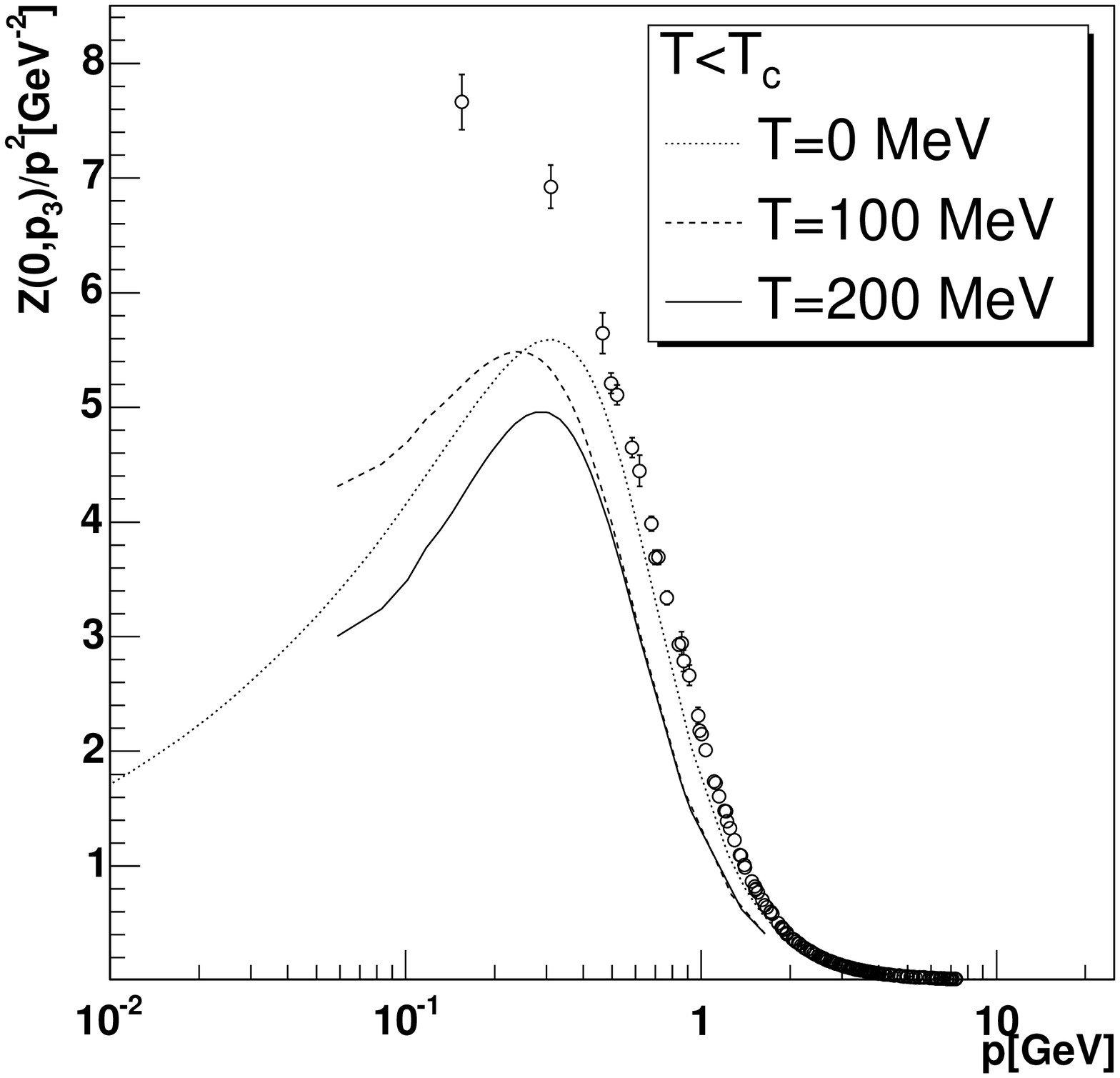, width=0.5\linewidth}\epsfig{file=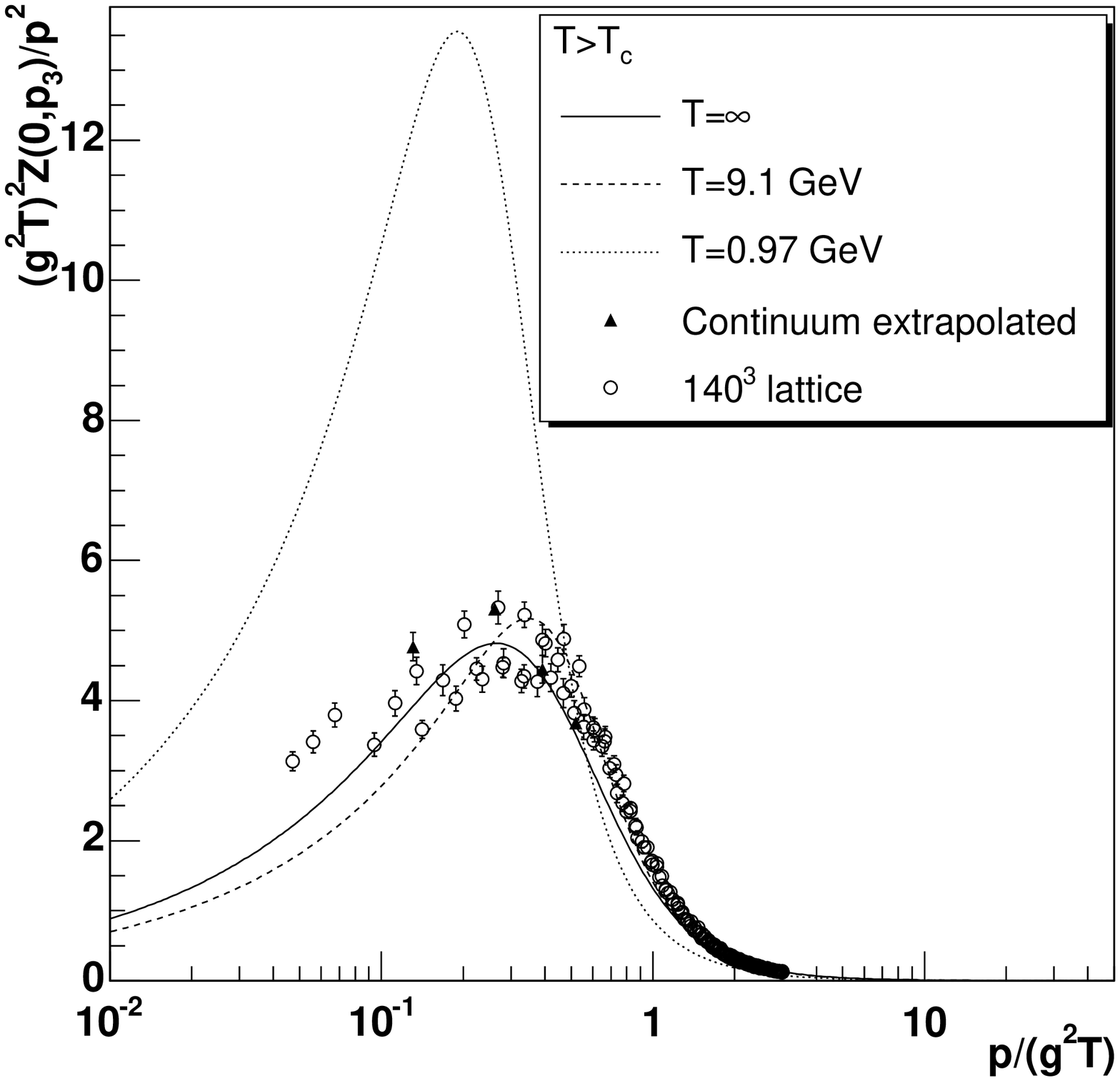, width=0.5\linewidth}
\caption{The transverse gluon propagator is shown in the vacuum\protect\cite{Fischer:2002hn} and at temperatures below the phase transition\protect\cite{Gruter:2004kd} in the left panel and above the phase transition\protect\cite{Maas:2004se,Maas:2005hs} in the right panel. The lattice data in the left panel are at zero temperature from a $20^3\times 64$ lattice\protect\cite{Bowman:2004jm} and in the right panel at infinite temperature continuum extrapolated\protect\cite{Cucchieri:2001tw} and from large lattices\protect\cite{Cucchieri:2003di}.}\label{figz}
\end{figure*}

\begin{figure*}
\epsfig{file=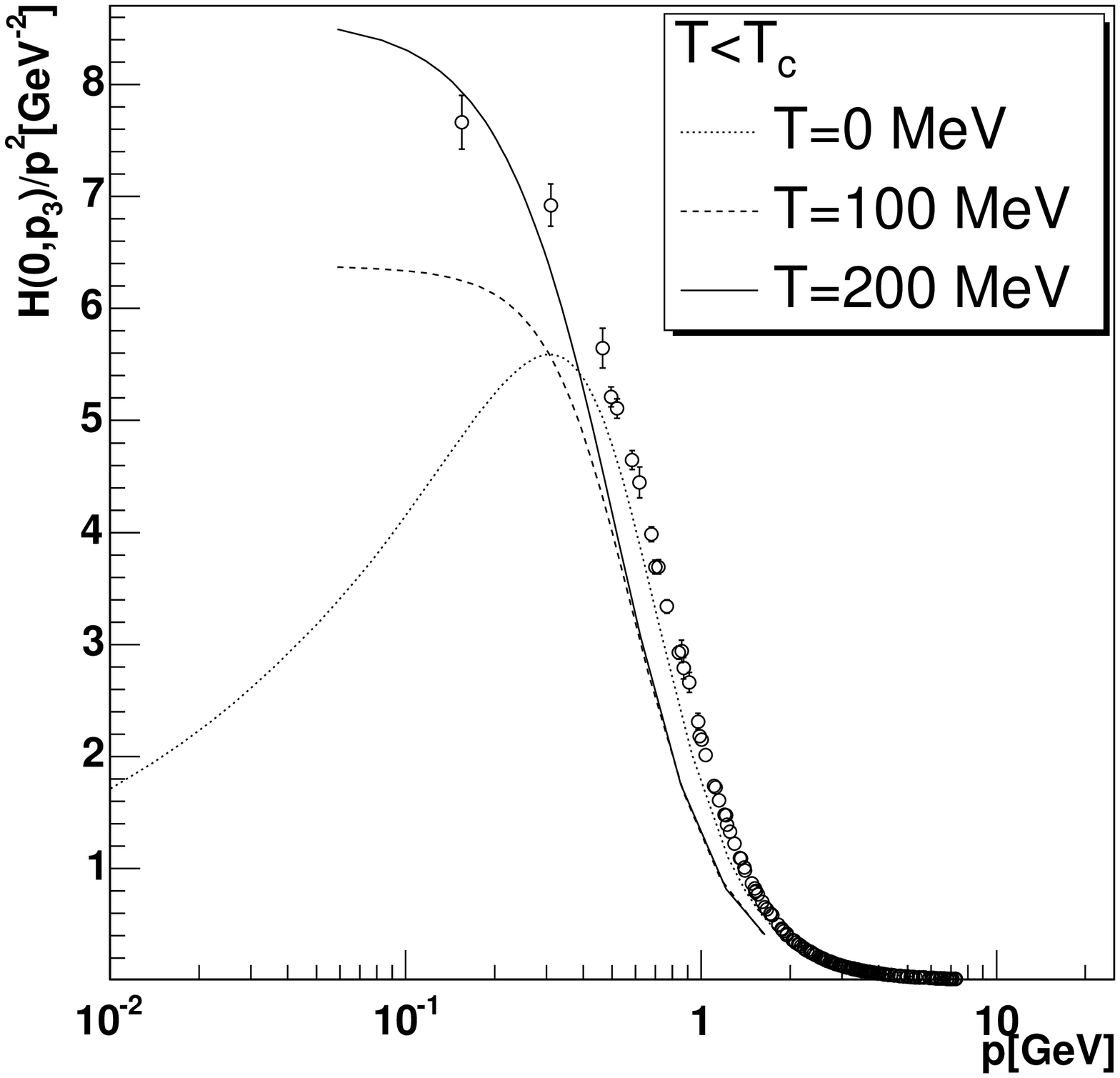, width=0.5\linewidth}\epsfig{file=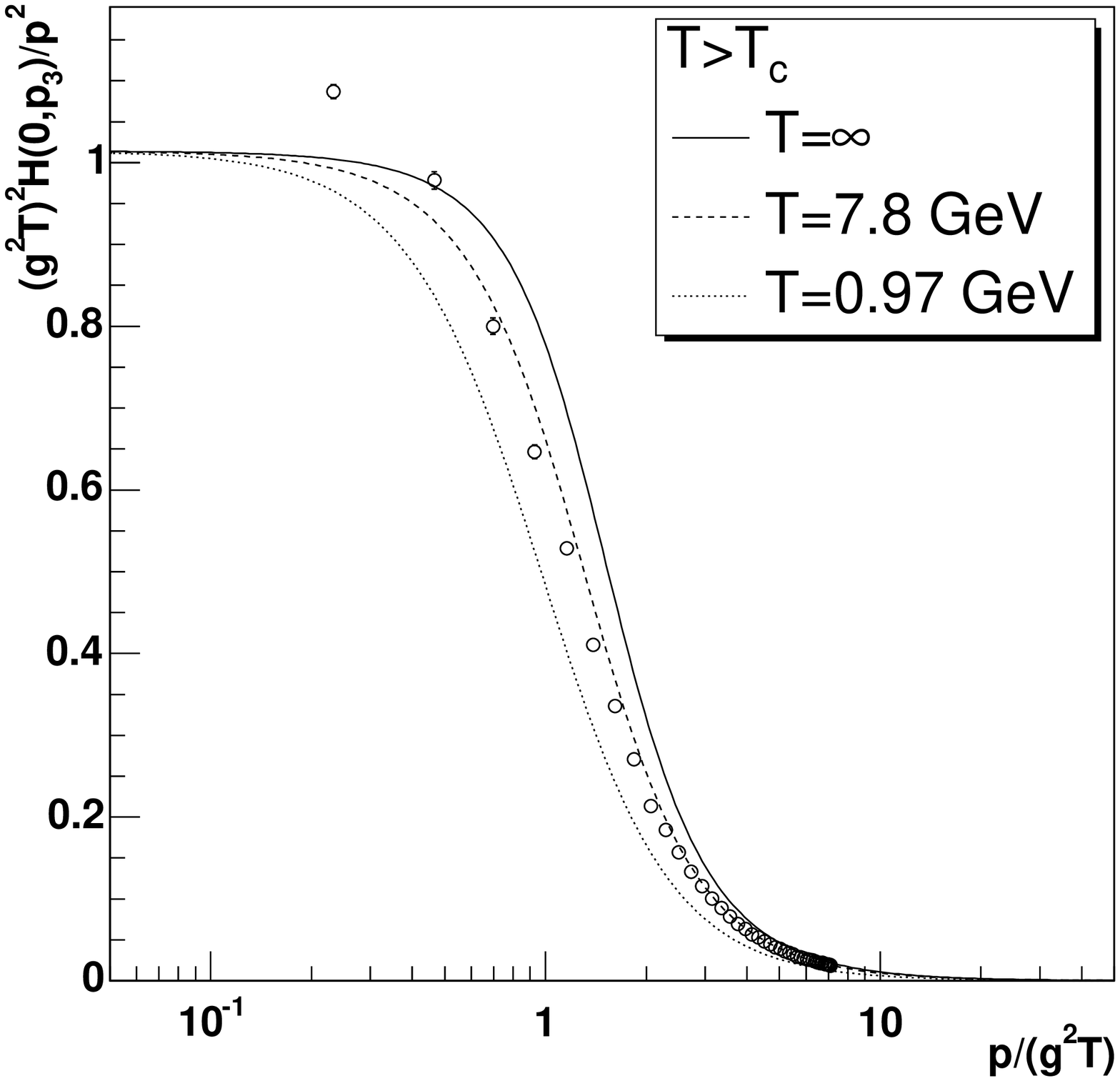, width=0.5\linewidth}
\caption{The longitudinal gluon propagator is shown in the vacuum\protect\cite{Fischer:2002hn} and at temperatures below the phase transition\protect\cite{Gruter:2004kd} in the left panel and above the phase transition\protect\cite{Maas:2004se,Maas:2005hs} in the right panel. The lattice data in the left panel are at zero temperature\protect\cite{Bowman:2004jm} and in the right panel at infinite temperature \protect\cite{Cucchieri:2001tw}.}\label{figh}
\end{figure*}

The results for the soft modes as a function of temperature are shown in figures \ref{figg}-\ref{figh}. For the high--temperature phase only the more likely solution is shown. The other is similar\cite{Maas:2004se,Maas:2005rf}, even quantitatively. The hard modes are essentially tree--level up to small corrections and are thus not shown here. Results for these can be found elsewhere for temperatures below\cite{Gruter:2004kd} and above\cite{Maas:2005hs,Maas:2005rf} the phase transition.

In figure \ref{figg} the ghost dressing function is displayed. The divergence in the infrared is clearly visible, satisfying \pref{Kugo}. The dressing function does not exhibit a marked qualitative dependence on temperature. Thus, the ghost is able to mediate long--range forces at all temperatures, in agreement with available lattice calculations.

The transverse gluon propagator is shown in figure \ref{figz}. It is infrared suppressed at all temperatures, satisfying \pref{Oehme}. Therefore transversely polarized gluons are always confined. Hence, the Yang--Mills theory remains non--trivial in the infinite--temperature limit. The peak at mid--momenta roughly signals the transition from perturbative to non--perturbative behavior. The height and to some extent the position of the peak is the region most sensitive to the truncation. The low--temperature propagators in both approaches do not fully reach into the infrared due to numerical limitations. In addition, the lattice data in the vacuum do not bend over due to finite volume effects. Only recent calculations on an extremely asymmetric lattice of $16^3\times 256$ find such a bending over\cite{Oliveira:2004gy}. Thus further investigations are of great interest. In the infinite--temperature limit, advantage can be taken of the lower dimensionality, and much larger lattices can be used. Such calculations show a clear bending over in the infrared\cite{Cucchieri:2003di}.

The situation is different for the longitudinal gluon propagator as shown in figure \ref{figh}. The low--temperature phase is confined, and the longitudinal and transverse gluon propagators are similar. Here, also the DSE results are not suppressed in the infrared. This is likely due to finite volume effects in the numerical method, as close investigations show\cite{Gruter:2004kd,Fischer:2005ui}. The weaker suppression in the infrared is more likely due to an increase of the coefficient $A_h$ in \pref{irh} rather than due to a change of the exponent $l$. More important is the behavior in the high--temperature phase. The propagator goes to a finite value in the infrared, in agreement with lattice results. Thus the propagator of gluons polarized longitudinally w.r.t.\ the heat--bath is essentially dominated by a (renormalized) mass.

The main difference between both phases is thus the longitudinal sector. This agrees with investigations of the temporal and spatial string tension\cite{Bali:1993tz}. The former corresponds to the longitudinal sector and vanishes in the high--temperature phase. The spatial string tension corresponds to the transverse sector and even increases. This also complies with a broken $Z_3$-symmetry in the high--temperature phase\cite{Holland:2000uj}, as the corresponding order parameter is only observed using time--like Polyakov lines. Furthermore, the lattice\cite{Cucchieri:2001tw,Nakamura:2003pu} and the DSE results\cite{Maas:2005hs,Maas:2005rf} suggests that the infinite--temperature limit is essentially reached already at temperatures of a few times $T_c$.

At this point the results can be compared to the expectations. Originally, the generation of chromoelectric and chromomagnetic screening masses has been expected. Long ago it was already shown\cite{Linde:1980ts} that a magnetic screening-mass can be at best of order $g^2T$ and not of perturbative origin, if it exists at all. This problem lead to the approach using HTLs\cite{Blaizot:2001nr}, which is based on resumming the hard mode contributions in self--energy diagrams, and then perform perturbation theory using these hard thermal loops. In the transverse infrared sector it is still plagued by severe problems, due to its perturbative nature, very similar to the vacuum case. In the high--temperature phase the HTL approach results in a transverse gluon propagator with a particle--like pole at $p=0$, thus $t=0$.

The DSE results disagree with these findings and yield that a magnetic screening mass does not exist and is not likely\cite{Maas:2004se,Maas:2005rf}. Instead over-screening persists as in the vacuum. This is in agreement with lattice calculations\cite{Nakamura:2003pu} at the highest available temperatures. HTLs and the DSEs find qualitatively similar results only for the propagators  of the soft longitudinal mode and the hard modes. However, the mass of the longitudinal mode is found\cite{Maas:2005hs,Maas:2005rf} to be of order $g^2T$ rather than of order $gT$, as expected in perturbative calculations.

\subsection{Positivity and confinement}\label{ssconfinement}

A sufficient criterion for confinement is the violation of positivity, which is already implied by the fulfillment of condition \pref{Oehme}. A more direct test is the determination of the corresponding Schwinger function, the Fourier transform of the propagators\cite{Alkofer:2003jk}. It is not positive semi--definite if positivity is violated. This has been found for the gluon propagator in the vacuum\cite{Alkofer:2003jk} and for the transverse gluon propagator in the high--temperature phase\cite{Maas:2004se,Maas:2005hs}, in agreement with lattice results\cite{Cucchieri:2001tw,Cucchieri:2004mf}.

As the longitudinal gluon propagator does not fulfill \pref{Oehme} in the high--temperature phase, it is of particular interest to determine its Schwinger function. It is found to be not positive semi--definite\cite{Maas:2004se,Maas:2005hs}. This result depends, however, much more strongly on the truncation than in the transverse sector. Thus, further investigations are mandatory to decide whether all gluons remain confined. The longitudinal gluon cannot be described  by leading--order perturbation theory alone, and non--perturbative or higher--order perturbative effects still contribute, as a comparison of lattice\cite{Cucchieri:2001tw} and DSE results shows\cite{Maas:2004se,Maas:2005rf}.

If the longitudinally polarized gluon is confined as well, this seems to occur through a different mechanism than for the transversely--polarized gluon. The properties of the Schwinger function indicate the possibility\cite{Maas:2005rf} for a Gribov--Stingl--like\cite{Habel:1990tw} confinement mechanism, but the results are not conclusive.

\subsection{A note on the thermodynamic potential}

Lattice calculations find a thermodynamic potential which seems to logarithmically approach a Stefan--Boltzmann behavior\cite{Karsch:2003jg,Karsch:2001cy,Boyd:1996bx}, but still differs significantly from it even at several times $T_c$. Besides lattice calculations, several attempts have been made to obtain the thermodynamic potential. Perturbation theory yields a slowly converging expansion of the potential, which differs from the ideal gas value even at very large temperatures\cite{Rischke:2003mt}. An alternative is based on perturbation theory in the hard and longitudinal modes and adding a few quantities from lattice simulations of the 3d Yang--Mills theory. This improved $g$ or weak-coupling expansion provides a good description of the lattice results\cite{Kajantie:2000iz}.

Using DSEs, the thermodynamic potential due to gluons can be obtained using the Luttinger--Ward/Cornwall--Jackiw--Tomboulis formalism\cite{Luttinger:1960ua}. It is, however, strongly limited by truncation artifacts. In the low--temperature phase, the thermodynamic potential is essentially constant, thus exhibiting the behavior expected of a confined system\cite{Gruter:2004kd}. At sufficiently high temperatures, the potential is dominated by the hard modes. These provide, up to small corrections, the potential of an ideal gas of gluons\cite{Kapusta:tk}, as the contribution due to the strong interactions is sub--leading. Hence despite the residual strong and confining interactions, a nearly trivial thermodynamic behavior is obtained. Near the phase transition, however, the contribution of the soft modes are likely to be important\cite{Zwanziger:2004np}.

In principle, knowledge of the thermodynamic potential gives access to the critical temperature, which is known from lattice calculation\cite{Karsch:2003jg} with high accuracy to be $T_c=269\pm 1$ MeV. Using DSEs it is not yet possible to extract the critical temperature. Nonetheless, observations which can be interpreted as super--heating\cite{Gruter:2004kd} and super--cooling\cite{Maas:2005hs,Maas:2005rf} together with the qualitative change in the longitudinal sector suggest a first--order phase transition. This would be correct for SU(N) with $N>2$, but not for other gauge groups\cite{Holland:2003mc}, {\it e.g.}\ SU(2).

\section{Discussion and conclusions}\label{sdiscuss}

Before discussing the implications of the results, it is important to assess their reliability. In the ultraviolet, resummed perturbation theory provides a reliable description due to asymptotic freedom. Thus, the incorporation of the results of perturbation theory is required for a final understanding. The infrared regime does not permit such a systematic approach. However, substantial evidence exists for the validity of the truncation scheme used in the DSE analysis\cite{Zwanziger:2003cf,Taylor:ff,Cucchieri:2004sq,Schleifenbaum:2004id,Alkofer:2004it}. These results are reinforced by the comparison with available calculations using lattice and RG methods. The gap at intermediate momenta is well under control using lattice calculations, which agree qualitatively with DSE calculations. Thus the presence of gluon confinement in the vacuum and in the infinite--temperature limit is a robust statement. This necessarily implies the non--triviality of the high--temperature phase of Yang--Mills theory, as expected.

At finite temperatures, and especially in the domain of the phase transition, the results for the propagators become less reliable. Both, DSE and lattice methods, have problems to systematically treat this range. For DSEs, the problem is present mainly due to the necessity to handle an infinite number of Matsubara frequencies, while lattice calculations suffer from the required volumes to reach small temperatures. This also entails that the extension in the time direction becomes too small to reliably extract the longitudinal propagator. Nonetheless, the lattice provides excellent results on thermodynamic properties. On the other hand, some general results on the infrared regime obtained from DSE calculations still hold. Thus qualitative results seem to be trustworthy. This is mainly possible due to the large effective `mass' of the hard modes, which is at the phase transition already of the order of 1.7 GeV. 

Still, much work has to be done to obtain the quantitative properties, including the structure of the phases. In addition, it is as yet unclear which are the correct colorless and thus gauge--invariant excitations governing the high--temperature phase. The persistence of confinement requires that this cannot be simple gluons, but perhaps light glueballs formed of soft gluons. These would likely not contribute significantly to thermodynamic quantities at sufficiently high temperatures. How this can be realized is subject to further investigations.

Nonetheless, a coherent picture emerges. The main difference between the low--temperature and the high--temperature phase is not primarily one between a strongly interacting and confining system and one with only quasi--free particles. Over--screening remains in at least part of the spectrum and thus confinement persists. In the transverse sector, the results are in accordance with the Gribov--Zwanziger and/or the Kugo--Ojima scenarios. The scenario that emerges is a chromoelectric phase transition of first order. In the vicinity of the phase transition, the non--perturbative effects are likely relevant to thermodynamic properties\cite{Zwanziger:2004np}, underlining the importance for experiments.

It is still unknown which effects can be induced by quarks. In the infinite--temperature limit, arguments exist\cite{Appelquist:vg} that they will not affect the propagators due to their fermionic nature. Fermions require anti--periodic boundary conditions\cite{Kapusta:tk}, and therefore generate effective `masses' of $(2n+1)\pi T$. Thus all quarks become infinitely massive in the infinite--temperature limit, irrespective of their intrinsic mass and thus decouple due to asymptotic freedom. At the phase transition, however, lattice calculations indicate\cite{Karsch:2003jg} that the quarks possibly are able to change the nature of the phase transition to a second order one or to a cross--over.

In addition, the observation of a drastic change in the inter--quark potential\cite{Kaczmarek:2004gv} and its connection to the residual confinement of gluons still needs to be understood. It cannot as yet be firmly concluded that the inter--quark potential still rises slowly at very large distances even at high temperatures. On the other hand, quark confinement is still not yet understood at zero temperature\cite{Alkofer:2003jk}, thus posing a more significant problem.

In this review the main results from studies on the finite--temperature behavior of Yang--Mills theories in Landau gauge have been summarized. The investigation is based primarily on the infrared properties of the propagators of the elementary degrees of freedom, the gluons and the Faddeev--Popov ghosts.

The results provide evidence for the existence of a phase transition in the pure gauge sector. In both phases strong interactions and non--perturbative features like confinement are found. Direct comparison to lattice and RG results in the vacuum and at high temperatures support this. However, many unsettled questions remain, such as the nature of the observable degrees of freedom in the high--temperature phase and the influence of quarks. Thus further investigations are desirable and in progress. These are accompanied by corresponding investigations at finite density\cite{Rischke:2003mt,Nickel:2005}. This embeds these findings in the ongoing effort not only to understand the vacuum structure of QCD but also to map out the phase diagram and to finally understand the results obtained in experiment from first principles.

\section*{Acknowledgments}

The author is grateful for critical readings and helpful remarks by Jochen Wambach and Reinhard Alkofer. This work was supported by the Helmholtz association (Virtual Theory Institute VH--VI--041).

\end{document}